\documentclass[letterpaper,titlepage,11pt]{article}
\usepackage{hyperref}
\usepackage{amssymb,amsmath,amsfonts,enumerate}
\usepackage{graphicx}

\def\clock{{\count0=\time
           \divide\count0 60
           \ifnum\count0<10 0\fi\the\count0
           \multiply\count0 -60 \advance\count0 \time
           :\ifnum\count0<10 0\fi \the\count0
         }}
\newcommand{\timestamp}{{\small\vbox{\hbox{\tt\jobname.tex}
\hbox{\the\day/\the\month/\the\year, \clock}}}}

\newcommand{\beq}{\begin{equation}}
\newcommand{\eeq}{\end{equation}}
\newcommand{\ben}{\begin{displaymath}}
\newcommand{\een}{\end{displaymath}}
\newcommand{\beqa}{\begin{eqnarray}}
\newcommand{\eeqa}{\end{eqnarray}}
\newcommand{\bea}{\begin{eqnarray}}
\newcommand{\eea}{\end{eqnarray}}
\newcommand{\bean}{\begin{eqnarray*}}
\newcommand{\eean}{\end{eqnarray*}}
\newcommand{\ba}{\begin{array}}
\newcommand{\ea}{\end{array}}
\newcommand{\bi}{\begin{itemize}}
\newcommand{\ei}{\end{itemize}}
\newcommand{\ie}{{\it i.e.,\,}}
\newcommand{\eg}{{\it e.g.,\,}}

\newcommand{\ka}{\kappa}

\numberwithin{equation}{section}

\begin{document}

\begin{titlepage}
\begin{flushright}
DCPT-10/31
\end{flushright}
\vskip 2.cm
\begin{center}
{\bf\LARGE{Multi-black rings and the phase diagram of higher-dimensional black holes}}
\vskip 1.5cm
{\bf Roberto Emparan$^{a,b}$, Pau Figueras$^{c}$
}
\vskip 0.5cm
\medskip
\textit{$^{a}$Instituci\'o Catalana de Recerca i Estudis
Avan\c cats (ICREA)}\\
\textit{Passeig Llu\'{\i}s Companys 23, E-08010 Barcelona, Spain}
\\
\textit{$^{b}$Departament de F{\'\i}sica Fonamental and}\\
\textit{Institut de
Ci\`encies del Cosmos, Universitat de
Barcelona, }\\
\textit{Mart\'{\i} i Franqu\`es 1, E-08028 Barcelona, Spain}\smallskip
\\
\textit{$^{c}$Centre for Particle Theory \& Department of Mathematical Sciences,}\\
\textit{Science Laboratories, South Road, Durham DH1 3LE, United Kingdom}\\
\vskip .2 in
\texttt{emparan@ub.edu, pau.figueras@durham.ac.uk}

\end{center}

\vskip 0.3in

\baselineskip 16pt
\date{}

\begin{center} {\bf Abstract} \end{center} 

\vskip 0.2cm 

Configurations of multiple concentric black rings play an important role
in determining the pattern of branchings, connections and mergers
between different phases of higher-dimensional black holes. We examine
them using both approximate and (in five dimensions) exact methods. By
identifying the role of the different scales in the system, we argue
that it is possible to have multiple black ring configurations in which
all the rings have equal temperature and angular velocity. This allows
us to correct and improve in a simple, natural manner, an earlier
proposal for the phase diagram of singly-rotating black holes in $D\geq
6$. 

\noindent

\end{titlepage} \vfill\eject

\setcounter{equation}{0}

\pagestyle{empty}
\small
\normalsize
\pagestyle{plain}
\setcounter{page}{1}

\newpage


\section{Introduction}

As a result of recent progress, a complete understanding of the possible
phases of some classes of higher-dimensional black holes appears now as
a realistic objective (some, but not all, of this progress, is reviewed
in \cite{Emparan:2008eg}). Here we focus on the most basic sector of
vacuum, asymptotically flat black holes. In five dimensions, there
appears to be little room for new solutions besides the basic ones of
the Myers-Perry black holes \cite{Myers:1986un}, black rings
\cite{Emparan:2001wn,Pomeransky:2006bd}, and helical black rings
\cite{Emparan:2009vd}, and combinations of all these
\cite{Elvang:2007rd,Iguchi:2007is,Evslin:2007fv,Izumi:2007qx,
Elvang:2007hs}.

The situation in $D\geq 6$, although much less developed than in five
dimensions, has also seen significant advances. In this paper we
focus on the sector of configurations with rotation in a single plane.
This is simple enough that we may conceivably narrow down the spectrum
of possibilities until only some details of branches and connections
remain to be determined, possibly through numerical work\footnote{For
the purposes of this paper, a `branch of solutions' means a
one-parameter family of black holes (possibly with multiple,
disconnected horizons) with fixed total mass and varying total angular
momentum. It is adequately represented
as a curve of area vs.\ angular momentum, at fixed mass.}. In this
direction, ref.~\cite{Emparan:2007wm} identified a large portion of the possible phases
and advanced proposals for the connections among them.

An important ingredient in the phase diagram are configurations
involving several concentric black rings. These are particularly
relevant for understanding branchings and connections among phases. It
was observed in \cite{Emparan:2003sy} that at sufficiently large
rotation, the class of Myers-Perry black holes should branch off
into new phases where the horizon develops axisymmetric `pinches' (this
has been subsequently confirmed in \cite{Dias:2009iu}). As one moves
along one of the new branches of pinched black hole solutions, it is expected
that these pinches grow and eventually pinch down to a singularity on
the horizon. Continuing the evolution in phase space, the horizon
develops into a non-simply connected horizon or splits into disconnected
components. The simplest instance of this phenomenon is a black hole
developing a pinch along its axis of rotation and connecting to a black
ring. The next simplest case is a black hole with a circular
axisymmetric pinch giving rise to a black Saturn. This picture was
developed in \cite{Emparan:2007wm} by combining evidence from a variety
of strands.

In the evolution in phase space of a configuration with a connected
horizon (I) that develops into a configuration (II) with two
disconnected black holes through an intermediate singular solution (I'),
it is important to realize that, since these are equilibrium
configurations, the temperature $T$ (\ie surface gravity) and angular
velocity $\Omega_H$ of the horizon must remain uniform on the horizon of
phase (I) and by continuity in phase (I') too\footnote{Note that we are
not demanding that $T$ and $\Omega_H$ remain constant as we move from
one solution to another in the same branch, but rather that they are
uniform over the horizon of each solution.}. It is then natural to
expect that the evolution can continue into a configuration (II) where
the two separate horizons still have equal temperatures and angular
velocities. This is, the branch of pinched black holes continues into a
branch of multiple black holes and black rings which are in
\textit{thermodynamic equilibrium}.

There is nothing wrong with a classical stationary configuration of
multiple black holes where each horizon has a different surface gravity
and angular velocity. However, if quantum effects are turned on, then
Hawking radiation will put these black holes in contact with each other
and the radiation will tend to equilibrate both their temperatures and
their angular velocities. Of course, here we are glossing over the fact
that the Hartle-Hawking state of thermodynamic equilibrium typically
does not exist for rotating black holes due to the presence of
ergoregions \cite{Kay:1988mu}. But this is not an issue for our
motivation to impose that the surface gravities and angular velocities
remain uniform in multi-horizon configurations
across the connection between two phases. This has nothing to do with
actual temperature and radiation, but, given the obvious thermodynamic
analogies, we will continue to refer to the configurations of our
interest as being in thermodynamic equilibrium.

Rotating black holes that develop several concentric pinches should
naturally evolve into configurations with several concentric black rings
in thermodynamic equilibrium (possibly with a central black hole of
spherical topology). However, ref.~\cite{Emparan:2007wm} found an
apparent objection against the existence of multi-rings in thermodynamic
equilibrium, which made it difficult to complete the phase diagram. The
main purpose of this paper is to revisit the properties of multi-ring
configurations, identify the loophole in the arguments in
\cite{Emparan:2007wm}, and then propose a new, simpler form for the
phase diagram of singly-rotating black holes in $D\geq 6$. The arguments
can be put to the test by studying the five-dimensional exact di-ring
solutions constructed in \cite{Iguchi:2007is} (see also
\cite{Evslin:2007fv}). Even if the details of the connections differ in
five dimensions (where all the mergers occur on a horizonless singular
solution) and in six or more dimensions, the properties of multi-rings
at very large angular momentum remain qualitatively similar.

In the next section we study multi-black rings in the approximation
where they are thin. Such methods, more generally developed in \cite{Emparan:2009cs},
had already been employed in an analysis of five-dimensional black
Saturns in \cite{Elvang:2007hg}, and give us the necessary understanding of the
physics of these configurations. In section \ref{sec:exact5d} we check
the validity of these methods against the exact five-dimensional black
di-rings. We also obtain the properties of di-rings at all angular
momenta, which is not possible within the thin-ring approximation.
Section~\ref{sec:newphase} gathers together the new information to draw
what we believe is a qualitatively correct phase diagram. We close in
Section~\ref{sec:fin} with some additional remarks. The appendix contains
details of the properties of the exact five-dimensional di-ring.

\medskip

\textit{Note:} when this project was in its final stages we learned from
T.~Mishima of his work with H.~Iguchi that overlaps with
section~\ref{sec:exact5d}.


\section{Multi-black rings in thermodynamic equilibrium}
\label{sec:multithin}

\subsection{Thin-ring approximation}

We say that a black ring in $D$ dimensions, with horizon topology
$S^1\times S^{D-3}$, is thin when the $S^1$ radius $R$ is much longer
than the $S^{D-3}$ radius $r_0$ that measures the ring `thickness'. In
this section we will work only to leading order in $r_0/R$, in which
case these two radii can be unambiguously defined. The physical
properties of black rings in this limit have been determined in
\cite{Emparan:2007wm} as
\beqa
M&\to&\frac{\Omega_{(D-3)}(D-2)}{8}R r_0^{D-4}\,,\qquad
J\to\frac{\Omega_{(D-3)}\sqrt{D-3}}{8}R^2 r_0^{D-4}\,,\nonumber\\
A_H&\to&2\pi\Omega_{(D-3)}\sqrt{\frac{D-3}{D-4}}R r_0^{D-3}\,,
\eeqa
\beq\label{TOm}
T\to\frac{D-4}{4\pi}\sqrt{\frac{D-4}{D-3}}\frac{1}{r_0}\,,
\qquad
\Omega_H\to\frac{1}{\sqrt{D-3}}\frac{1}{R}\,.
\eeq
(we set $G=1$). We introduce a dimensionless angular momentum and
area appropriate for
comparing different black objects with a given mass,
\beq
j=c_J\frac{J}{M^\frac{D-2}{D-3}}\,,\qquad
a_H=c_A\frac{A_H}{M^\frac{D-2}{D-3}}\,
\eeq
where the $D$-dependent numerical constants $c_J$ and $c_A$ have been chosen in
\cite{Emparan:2007wm,Emparan:2008eg} as
\beq
c_J =\frac{\Omega_{D-3}}{2^{D+1}}\frac{(D-2)^{D-2}}{(D-3)^{\frac{D-3}{2}}}\,,\qquad
c_A=\frac{\Omega_{D-3}}{2(16\pi)^{D-3}}(D-2)^{D-2}
\left(\frac{D-4}{D-3}\right)^{\frac{D-3}{2}}\,
\eeq
but these specific values are actually irrelevant to our arguments.
Thus, for a single thin black ring
\beq
j\to 2^{-\frac{D-2}{D-3}}\left(\frac{R}{r_0}\right)^{(D-4)/(D-3)}\,,
\eeq
and
\beq \label{aHj}
a_H^\mathrm{(single-ring)}\to  2^{\frac{D-
6}{(D-4)(D-3)}}j^{-1/(D-4)}\,.
\eeq
The thin ring limit is then
equivalent to a large $j$ limit. The separation of the length
scales $r_0\ll R$ implies that the thin ring can be regarded as a black string
curved on a circle of large radius. To leading order, the gravitational
interaction among segments of the string separated by a distance $\sim
R$ can be neglected.

\subsection{Multiple thin rings and thermal equilibrium}

The quantities $a_H$ and $j$ can also be introduced
for generic multi-black hole phases, where $M$, $J$ and $A_H$ are the
total mass, spin and area of the entire system. As argued in \cite{Elvang:2007hg},
for a system with $n$ black objects (\ie\ concentric black rings with or
without a central black hole) specifying the total $M$ and $J$ leaves $2n-2$ continuous
parameters undetermined.
In general these are configurations where 
the temperature and
angular velocities of the multiple horizons, $T^{(i)}$, $\Omega_H^{(i)}$
are different and so the system is out of thermodynamic equilibrium,
which is perfectly acceptable for a stationary classical black hole
system. However, as discussed in the introduction we will be particularly
interested in thermodynamic equilibrium where 
\beq
T^{(i)}=T^{(j)}\,,\qquad \Omega_H^{(i)}=\Omega_H^{(j)} \qquad \forall i, j\,.
\eeq
These are $2n-2$ conditions, which remove all the continuous
non-uniqueness. Thus multi-black hole phases in thermal
equilibrium correspond to curves $a_H(j)$.

The thin ring approximation can be used in order to
describe black Saturns in which the radius of the central black hole
$r_\mathrm{bh}$ is much smaller than the ring's $S^1$ radius $R$, so the
interaction
between the two objects is negligible and the total mass, spin, and
area are just the sum of the two separate components \cite{Elvang:2007hg}. It is easy
to deduce that for a thermally-equilibrated black Saturn in any number of
dimensions, the dimensionless total area $a_H(j)$ asymptotes (from below)
to the same value as for a single black ring 
\beq\label{arearatios}
\frac{a_H^\mathrm{(Saturn)}}{a_H^\mathrm{(single-
ring)}}\stackrel{j\to\infty}{\longrightarrow} 
1\,.
\eeq 
The reason is that equality of the temperatures forces the black hole
radius $r_\mathrm{bh}$ to be of the same order as the ring's $r_0$, but the
mass, spin and area of the ring are much larger since they contain
factors of $R\gg r_0$ so the black hole contributes negligibly to them.

We now extend these arguments to understand the properties
of multi-black ring systems. To begin with, let us consider two thin
black rings with radii $R^{(2)}<R^{(1)}$ and horizon thicknesses $r_0^{(1)}$, 
$r_0^{(2)}$. The interaction energy between them will be smaller than
their masses as long as 
\beq
r_0^{(1,2)}\ll R^{(1)}-R^{(2)}\,.
\eeq
Then we can approximate, to leading order in $r_0^{(i)}/R^{(i)}$ and in 
$r_0^{(i)}/(R^{(1)}-R^{(2)})$, the total mass and angular momentum
of the system as
\beq\label{totMJA}
M\simeq M^{(1)}+M^{(2)}\,,\qquad J\simeq J^{(1)}+J^{(2)}\,.
\eeq
The total area is always the sum of individual areas, $A_H=
A_H^{(1)}+A_H^{(2)}$.
Fixing $M$ and $J$ leaves two continuous parameters unspecified in the
di-ring configuration, which could be \eg the mass and spin of one of
the black rings. 

If we impose the two conditions that the temperatures
and angular velocities of the two black rings are equal, then the
doubly-continuous degeneracy is removed. 
However, since fixing $T$ fixes the thickness
$r_0$ and fixing $\Omega_H$ fixes the radius $R$ (see \eqref{TOm}), 
naively it would
appear that thermal equilibrium on thin black di-rings requires
that the two rings should actually be
the same, \ie they would lie on top of each other.
In other words, the interaction between them would be very strong so the initial
approximation breaks down, and
conceivably thermally-equilibrated di-rings might not be allowed \cite{Emparan:2007wm}.

What this argument misses is that it is possible to have a hierarchy
between the three main scales in the problem,
\beq\label{threescales}
r_0\ll R^{(1)}-R^{(2)}\ll R^{(i)}
\eeq
such that one can consistently neglect the interaction between the two
rings while at the same time satisfying the thermal equilibrium
conditions\footnote{Here $R^{(1)}$ and $R^{(2)}$ are of the same
parametric order, while $r_0^{(1)}$ and $r_0^{(2)}$ are equal to leading
order of approximation.}. 

The way to conceive of this configuration is by first considering two
parallel boosted black strings, with equal horizon thickness $r_0$ and
equal boost parameters. Then, the two black strings have the same
temperature and horizon velocity, and if the distance between them is
$L\gg r_0$ then we can linearly superimpose the solutions to a good
approximation (precisely, $O(r_0/L)^{D-3}$). Next, we bend this
di-string into a large circle of radius $R\gg L$ so the interaction of
segments of the di-string at distance $\sim R$ can be neglected. The
balance between the tension of the di-ring and the centrifugal force can
easily be seen to fix the boost value to the same parameter as in a
single ring, and we can obtain the physical properties as in
\eqref{totMJA}. Clearly, since $L=R^{(1)}-R^{(2)}$ this is the regime
\eqref{threescales}. In the language of the blackfold effective-theory
approach, we integrate successively the physics at two scales, first
$r_0$ (the thickness of the black strings), then $R^{(1)}-R^{(2)}$ (the
`thickness' of the di-ring).

Under these conditions, the mass, spin and area of the two black rings
(which in our approximation are well-defined quantities) are equal to
leading order in the approximation \eqref{threescales}, 
\beq
M^{(1)}\simeq M^{(2)}\,,\qquad J^{(1)}\simeq J^{(2)}\,,\qquad
A_H^{(1)}\simeq A_H^{(2)}\,,
\eeq
so for the entire di-ring system we
have
\beqa
j&=&c_J\frac{2J^{(1)}}{(2M^{(1)})^\frac{D-2}{D-3}}=2^{-1/(D-3)}j^{(1)}\,,
\\
a_H&=&c_A\frac{2A_H^{(1)}}{(2M^{(1)})^\frac{D-2}{D-3}}=2^{-1/(D-3)}a_H^{(1)}
\eeqa
where $j^{(1)}$ and $a_H^{(1)}$ are the values for each individual ring.
Since from \eqref{aHj} we know how $a_H^{(1)}$ depends on $j^{(1)}$, we can easily
determine the asymptotic form of $a_{H}(j)$ for a thermally-equilibrated di-ring
at large total angular momentum $j$. The result is that,
for a single black ring and a black di-ring with the same total mass and
spin, the dimensionless areas
asymptote to a finite, non-zero ratio,
\beq\label{arearatiodi}
\frac{a_H^\mathrm{(di-ring)}}{a_H^\mathrm{(single-ring)}}
\stackrel{j\to\infty}{\longrightarrow} 
2^{-1/(D-4)}\,.
\eeq

The argument can easily be extended to $n$ black ring configurations
in thermal equilibrium
when they satisfy the condition that
\beq
r_0\ll R^{(i)}-R^{(j)}\ll R^{(i)}
\eeq
for all pairs $R^{(i)}>R^{(j)}$ so their mutual interactions are negligible. Then
\beq\label{arearation}
\frac{a_H^\mathrm{(\mathit{n}-ring)}}{a_H^\mathrm{(single-
ring)}}\stackrel{j\to\infty}{\longrightarrow} 
n^{-1/(D-4)}\,.
\eeq

Finally, it is equally straightforward to argue that for a black Saturn
with $n$ black rings in thermodynamic equilibrium, $a_H$ is
asymptotically the same at large $j$ as for the $n$-ring without a
central black hole. This is, the asymptotic behavior of $a_H(j)$ only
depends on the number of rings in the system.


\section{Exact results from di-rings in five dimensions}
\label{sec:exact5d}

In five dimensions we are not constrained to the approximation where the
interaction between the black rings is small. Exact di-ring solutions
have been constructed in \cite{Iguchi:2007is,Evslin:2007fv} using
integrability techniques. In
the appendix we give the exact expressions for the total $M$, $J$, $A_H$
of the configuration, as well as the individual $A_H^{(i)}$, $T^{(i)}$,
$\Omega_H^{(i)}$, which are well-defined magnitudes on each horizon. In
five dimensions, the dimensionless magnitudes that we
use to describe the phase space at fixed mass are conventionally
normalized as
\begin{equation}
 a_H=\frac{3}{16}\sqrt{\frac{3}{\pi}}
\frac{A_H}{M^{3/2}}\,,\qquad
j=\sqrt{\frac{27\pi}{32}}\,\frac{J}{M^{3/2}}\,.
\end{equation}

As mentioned above, fixing $M$ and $J$ leaves
two parameters, \eg the mass and spin of one of the black rings,
unspecified. We can easily argue that this allows di-rings to take any value of
$j$ and $a_H$ within the
band 
\beq
-\infty<j<\infty\,,\qquad 0< a_H<1\,.
\label{theband}\eeq
The upper bound $a_H=1$ is
set by the maximum area that a single black ring can have. The argument
is similar to the one employed in \cite{Elvang:2007hg} for black Saturns: we can
consider a central ring with area arbitrarily close to $a_H=1$, and then
surround it with a long and thin ring that carries arbitrarily little
entropy and mass but which can contribute an arbitrary large amount of angular
momentum to the total configuration by making its radius as
long as required. The parameters can be adjusted to
reproduce any value of $j$ and $a_H$ within the band \eqref{theband}.
Indeed, typically there is a one-parameter family of di-rings at any given
point in \eqref{theband}.

\begin{figure}[t]
\begin{center}
 \includegraphics[scale=0.8]{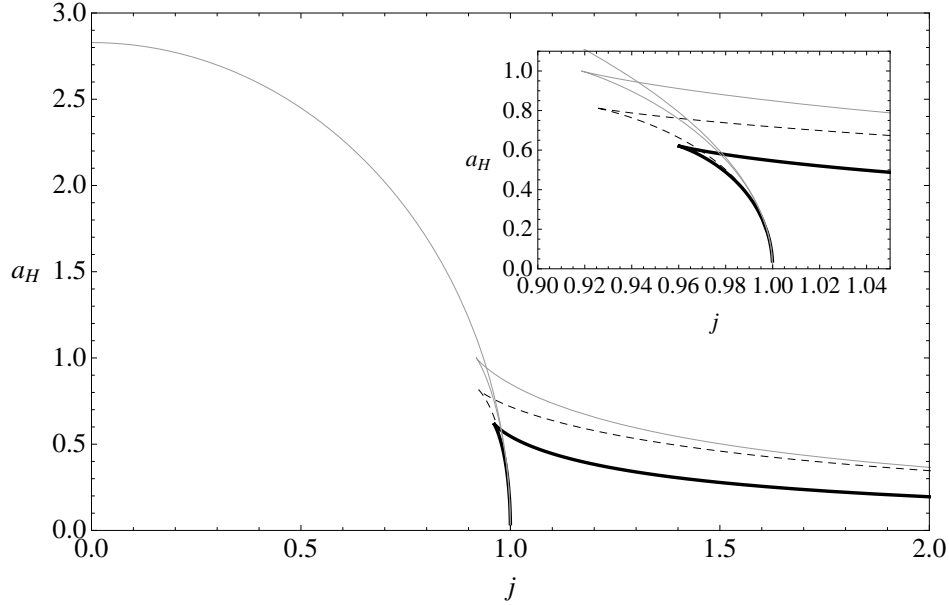}
\end{center}
\caption{\small Singly-rotating phases in thermodynamical equilibrium in
five dimensions. The gray curves correspond to the singly-spinning
Myers-Perry black hole and the black ring, the dashed curve corresponds to the
black Saturn and the black curve to the di-ring. For the black Saturn,
the maximum of the total area and minimum of the spin is at
$(a_\textrm{H}^{\textrm{(Saturn)}},j)\approx
(0.814, 0.9245)$
\cite{Elvang:2007rd}. For the di-ring it is at
$(a_\textrm{H}^{\textrm{(di-ring)}},j)\approx (0.623, 0.9596)$. The large-$j$
asymptotic behavior predicted in \eqref{arearatios}, \eqref{arearatiodi}
can be discerned already at $j\sim 2$.}
\label{fig:phases} 
\end{figure}

As before, the conditions of thermodynamical equilibrium  
\begin{equation}
\label{eqn:thermoeq}
 T^{(1)}=T^{(2)} \,,\qquad \Omega_H^{(1)}= \Omega_H^{(2)}\,
\end{equation}
remove completely the continuous non-uniqueness (leaving at most discrete
degeneracies) so these phases are characterized by curves $a_H(j)$. 
The resulting phase diagram is depicted in fig.~\ref{fig:phases}.

The most notable feature of this diagram is that the
different curves all join at the singular extremal black ring at $j=1$,
$a_H=0$, and that the area curve of black di-rings always lies below the
corresponding black ring curve. 
Let us focus on the large-$j$ asymptotic behavior, as
obtained from the explicit exact solutions. For a
single black ring the asymptotic value is (see \eg \cite{Emparan:2006mm})
\begin{equation}
 a_\textrm{H}^{\textrm{(single-ring)}}= \frac{1}{\sqrt{2}\,j}+O(j^{-3})\,,
\end{equation}
which is known to be reproduced exactly by the thin-ring (blackfold)
methods of \cite{Emparan:2007wm,Emparan:2009cs}.

Using the exact
results in the appendix, we can obtain the
asymptotic form of the di-ring curve 
in the limit of large $j$. We find\footnote{The first subleading terms are
$a_\textrm{H}=\frac{1}{2\sqrt{2}\,j}+\frac{3}{32\times2^{5/6}}\,
\frac{1}{j^{7/3}}+\frac{1}{16\sqrt 2}\,\frac{1}{j^3}+O(j^{-11/3})
$.}
\begin{equation}
\label{eqn:asyaH}
 a_\textrm{H}^{\textrm{(di-ring)}}=\frac{1}{2\sqrt{2}\,j}+O(j^{-\frac{7}{3}})\,.
\end{equation}
We see that our general argument that yielded the asymptotic
result \eqref{arearatiodi} indeed reproduces the correct value in $D=5$. 

We can further check that in this configuration the separation of scales
\eqref{threescales} occurs as $j\to\infty$. Using the exact solution of
\cite{Evslin:2007fv}, we define\footnote{Thus we take $R$
as the radius of the $S^1$ at the outer rim of the black ring horizon, and $r_0$ as the
radius of the $S^2$ at the (coordinate) mid-point between the
poles of the horizon $S^2$ of the outer black ring.}
\begin{equation}
 R^{(1),(2)}=\sqrt{g_{\psi\psi}}\,\bigg|_{\rho=0,\,z=a_{2,5}}\,,
\qquad r_0=\sqrt{g_{\phi\phi}}\,\bigg|_{\rho=0,\,z=\frac{a_2+a_3}{2}}\,.
\end{equation}
In the limit of large $j$ the difference between alternative definitions
of these quantities (see \cite{Elvang:2006dd}) vanishes as a positive
power of $1/j$. From the exact
solution we obtain
\begin{equation}
\begin{aligned}
\frac{r_0}{R^{(1)}-R^{(2)}}&=\frac{1}{8 \times2^{5/6}}\frac{1}{j^{4/3}}
-\frac{1}{64 \sqrt{2}\, j^2}+O(j^{-8/3})\,,\\
\frac{R^{(1)}-R^{(2)}}{R^{(1)}}&=\frac{1}{2^{2/3}}\frac{1}{j^{2/3}}
-\frac{1}{4\times 2^{1/3}}\frac{1}{j^{4/3}}+\frac{3}{16\, j^2}+O(j^{-8/3})\,,
\end{aligned}
\end{equation}
so we are indeed in the regime \eqref{threescales}.
Thus our assumption that at large $j$ the di-ring exists in a thermodynamic
equilibrium configuration where the interactions between rings are
weak, is well-supported by the exact solution.


\section{Phase diagram revisited}
\label{sec:newphase}

It was conjectured in \cite{Emparan:2003sy}, and recently confirmed in
\cite{Dias:2009iu}, that for a sufficiently large spin, singly-rotating
black holes in $D\geq 6$ branch off into new phases of stationary black
holes with axisymmetric `pinched' horizons. 

It is expected that moving in the space of solutions along a new branch
of stationary black holes, the deformation increases until the horizon
pinches down to zero size at the axis of rotation or at a circle
centered on the axis. The simplest instance corresponds to the merger of
black holes with a single pinch along the rotation axis, with `fat' black
rings whose central hole has shrunk to zero size. Similarly, black holes
with one circular pinch are expected to merge with black Saturns. An
important point here is that, if we follow the evolution in phase space
of these solutions, then coming from the side of the pinched black hole
the temperature and angular velocity must be uniform on the horizon. By
continuity, this will also be the case at the merger point and then it
is natural to expect that the phases branch off into configurations of black
Saturns in thermal equilibrium. That is, thermal equilibrium phases of
multi-black holes are important for determining the phase structure near
merger points.

Using this information, ref.~\cite{Emparan:2007wm} made a proposal for the phase
diagram of singly-spinning black objects in $D\geq 6$. Some of the
qualitative structure of the conjectured diagram is robust ---these are
the solid lines and figures in fig.~6 of ref.~\cite{Emparan:2007wm}. However, there
were some phases
and connections that remained more uncertain---these are the dashed lines
and figures in fig.~6 of ref.~\cite{Emparan:2007wm}. They were motivated by the
apparent absence of multi-rings in thermal equilibrium, which we can
now assert is incorrect.

\begin{figure}[th!]
\centerline{\includegraphics[width=12cm]{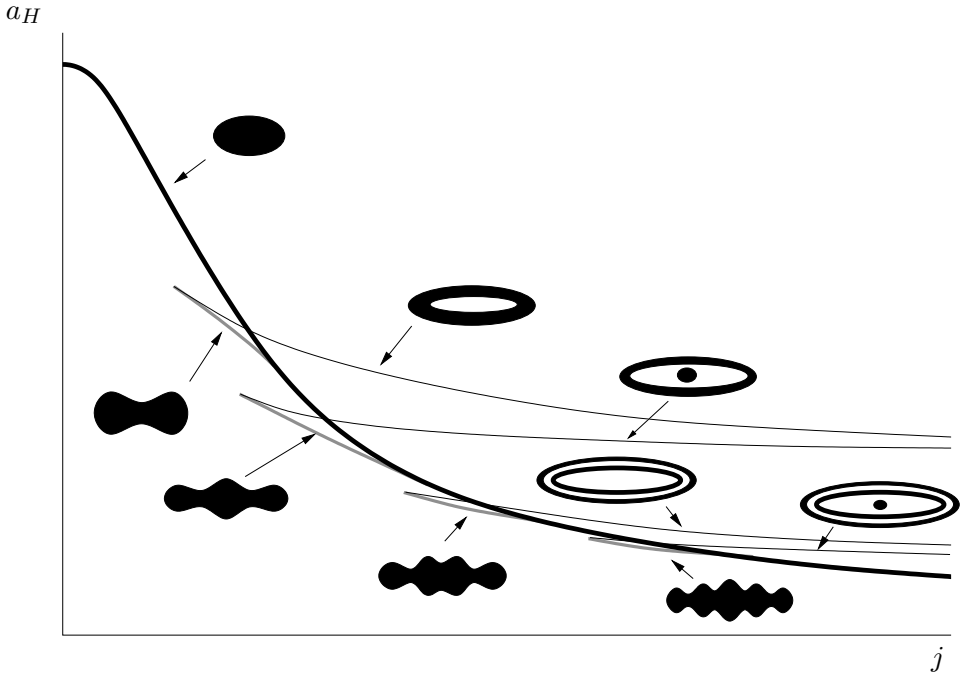}}
\begin{picture}(0,0)(0,0)
\put(30,250){$a_H$}
\put(380,5){$j$}
\end{picture}
\caption{\small New proposal for the phase diagram of thermal equilibrium phases
in $D\geq 6$. As in ref.~\cite{Emparan:2007wm}, the details of the phase connections are unknown
and smooth connections (\ie second order transitions)
are possible instead of swallowtails with cusps (\ie first order
transitions). In phases of black holes with multiple pinches evolving
into multi-black rings (and multi-ring Saturns) it is also
unknown whether intermediate pinched Saturns or pinched multi-rings
appear (this depends on how the different pinches evolve along the phase
curve).
Other than this, the features in the diagram are robust. The
asymptotic behavior of the curves depends only on
the total number of rings and is given by eq.~\eqref{arearation}.}
\label{fig:hidphases}
\end{figure}

A first amendment to the picture in \cite{Emparan:2007wm} has been
already made in ref.~\cite{Emparan:2009vd}. Using the blackfold approach
it has been argued that the `pancaked black Saturns' conjectured in
\cite{Emparan:2007wm} cannot exist in thermal equilibrium: an
ultraspinning black hole has larger radius than a black ring with the
same temperature and angular velocity, and therefore would engulf it.
Fortunately, our analysis in this paper shows that these pancaked
Saturns are not actually needed to complete the phase diagram and
describe the mergers of black holes with several pinches. Instead these
are naturally connected to multi-black rings in thermal equilibrium. The
corrected phase diagram is presented in fig.~\ref{fig:hidphases}.

It must be noted that we cannot decide at present whether two pinches in
a black hole will shrink to zero size simultaneously as one moves along
the curve $a_H(j)$ in the phase diagram. It may well happen that one of
the pinches reaches zero size before the other, thus giving rise to a
phase where a pinched black hole is surrounded by a black ring, or
instead one where a pinched black ring surrounds a central black hole.
Ascertaining this probably requires explicit numerical construction of
the solutions. Up to these uncertainties about the details of how the
different phases actually connect, we believe that the qualitative
features presented in fig.~\ref{fig:hidphases} are robust.


\section{Final remarks}
\label{sec:fin}

An incorrect assumption about the properties of multi-black rings in
thermal equilibrium had led ref.~\cite{Emparan:2007wm} to conjecture an
unduly complicated phase diagram in $D\geq 6$. In this
paper, after properly identifying the possible hierarchies between the
length scales in the system, we have concluded that the simplest and
most natural completion of the diagram can actually be realized. An
analysis of the exact five-dimensional di-ring solutions confirms in
detail the results obtained by performing a thin-ring
approximation.

While in $D\geq 6$ we have worked only to leading order in the thin-ring
approximation, it should not be too difficult to estimate the size of
subleading corrections in
$r_0/(R^{(1)}-R^{(2)})$  by considering a linearized
(Newtonian) approximation to the gravitational interaction between black
objects.

We have only studied configurations with a single angular
momentum. But arguments similar to the ones in this paper can be made
for other multi-black hole systems, where instead of singly-spinning
black rings we have doubly-spinning black rings \cite{Pomeransky:2006bd}, helical black
rings, or blackfolds with horizon topology $\prod_{i| p_i\in
\mathrm{odd}}S^{p_i}\times S^{D-\sum_i p_i-2}$ \cite{Emparan:2009vd}.

Observe that the five-dimensional black Saturn and di-ring phases merge
with other black hole phases (MP black hole, black ring) at the singular
extremal solution at $j=1$, $a_H=0$. Presumably, this is also the case
for all multi-ring phases, and it suggests that bifurcations of
horizons (at least singly-rotating ones) do not occur in five
dimensions. Indeed, it has been proven that singly-rotating MP black
holes in $D=5$ do not bifurcate \cite{Dias:2010maa}, and this is likely
the case for black rings too.


\section*{Acknowledgments}

RE thanks Takashi Mishima for conversations and communications about
their work overlapping with ours. PF thanks Oscar J.C. Dias, Ricardo Monteiro and Jorge E. Santos for many discussions about di-rings in thermodynamic equilibrium. RE thanks Yukawa Institute, Kyoto
University (during the long-term workshop ``Gravity and Cosmology
2010'') and KIAS (Seoul) for hospitality during the last phases of this
work. RE is partially supported by DURSI 2009 SGR 168, MEC FPA
2007-66665-C02 and CPAN CSD2007-00042 Consolider-Ingenio 2010. PF is supported by an STFC rolling grant.


\appendix

\section{The black di-ring} \label{sec:details} 
\begin{figure}[t]
\vspace{2cm}
\begin{center}
\begin{picture}(-1,0)
\setlength{\unitlength}{1cm}
\put(-5,1.5){\line(1,0){11}}
\put(-2.5,1.5){\linethickness{0.1cm}{\line(1,0){1.5}}}
\put(1.,1.5){\linethickness{0.1cm}{\line(1,0){1.5}}}
\put(-5,0.5){\line(1,0){11}}
\put(-5,0.5){\linethickness{0.1cm}{\line(1,0){2.5}}}
\put(-1.0,0.5){\linethickness{0.1cm}{\line(1,0){2.}}}
\put(2.5,0.5){\linethickness{0.1cm}{\line(1,0){1.5}}}
\put(-5,-0.5){\line(1,0){11}}
\put(4,-0.5){\linethickness{0.1cm}{\line(1,0){2}}}
\put(-5.5,1.5){$t$}
\put(-5.5,-0.5){$\psi$}
\put(-5.5,0.5){$\phi$}
\put(-2.65,1.75){\small{$(1,0,\Omega_H^{(1)})$}}
\put(0.9,1.75){\small{$(1,0,\Omega_H^{(2)})$}}
\put(-4.5,0.75){\small{$(0,1,0)$}}
\put(-0.6,0.75){\small{$(0,1,0)$}}
\put(2.65,0.75){\small{$(0,1,0)$}}
\put(4.3,-0.25){\small{$(0,0,1)$}}
\put(-3.8,-1.1){$0$}
\put(-2.5,-1.1){$\ka_1$}
\put(-1.1,-1.1){$\ka_2$}
\put(0,-1.1){$\ka_3$}
\put(1.,-1.1){$\ka_4$}
\put(2.5,-1.1){$\ka_5$}
\put(3.9,-1.1){$1$}
\end{picture}
\end{center}
\vspace{0.5cm}
\caption{\small{Rod structure of the di-ring.}}
\label{fig:diring}
\end{figure}
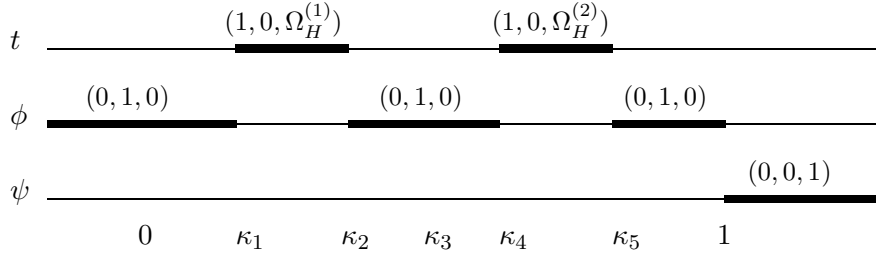

In this appendix we
collect some of the physical parameters of the black di-ring that are
needed in our analysis. The solution was first presented in \cite{Iguchi:2007is} but
we follow mostly the inverse-scattering construction of
\cite{Evslin:2007fv}, to which we refer for more details. The rod
structure is as in fig.~\ref{fig:diring}, and the solution
is parametrized by seven real numbers
$a_i$ with dimensions of length$^2$ that correspond to the
rod endpoints and satisfy
\begin{equation}
 a_1<a_2<a_3<a_4<a_5<a_6<a_7\,.
\end{equation}
In this parametrization the black ring horizons lie at $\rho=0$ and
$z\in [a_2,\,a_3]$ and $z\in [a_5,\,a_6]$ respectively, where $\rho$ and
$z$ are the standard Weyl coordinates. 

The ADM mass and angular momentum of the solution are \cite{Evslin:2007fv}:
\begin{equation}
\begin{aligned}
&M=\frac{3\,\pi}{4}\left(a_3-a_1+a_6-a_4\right)\,,\\
&J=\frac{\pi}{2}\frac{(a_2-a_1)(a_5-a_1)\,c_1+(a_4-a_2)(a_5-a_4)c_2}{a_4-a_1}\,,
\end{aligned}
\end{equation}
where $c_1$ and $c_2$ are constants which are fixed by demanding
regularity of the solution: 
\begin{equation}
\label{eqn:thecs}
c_1=\sqrt{\frac{2(a_3-a_1)(a_6-a_1)(a_7-a_1)}{(a_2-a_1)(a_5-a_1)}}\,,\qquad
c_2=\sqrt{\frac{2(a_4-a_3)(a_6-a_4)(a_7-a_4)}{(a_4-a_2)(a_5-a_4)}}\,.
\end{equation}

Once $c_1$ and $c_2$ have been fixed as in \eqref{eqn:thecs}, the
absence of conical singularities imposes two further constraints on the
parameters of the solution \cite{Evslin:2007fv}:
\begin{equation}
\label{eqn:conicalsingus}
\frac{|Y-c_1\,c_2\,Z|}{\sqrt{X}}=1\,,\qquad
\frac{(a_7-a_1)(a_7-a_4)(a_7-a_3)(a_7-a_6)}{(a_7-a_2)^2(a_7-a_5)^2}=1\,,
\end{equation}
where
\begin{equation}
\begin{aligned}
&X=\frac{4(a_4-a_1)^2(a_5-a_2)^2(a_6-a_3)^2(a_7-a_2)^2(a_4-a_3)(a_6-a_1)(a_7-a_1)}{(a_4-a_2)(a_5-a_1)(a_5-a_3)(a_6-a_2)(a_7-a_3)}\,,\\ 
&Y=2(a_4-a_3)(a_6-a_1)(a_7-a_1)\,,\\
&Z=(a_2-a_1)(a_5-a_4)\,.
\end{aligned}
\end{equation}
The conditions in \eqref{eqn:conicalsingus} imply that the general
(regular) di-ring is described by an overall length scale and three
dimensionless parameters \cite{Evslin:2007fv}. As discussed in the main
text, this number of parameters is to be expected since a black ring in
equilibrium is uniquely specified by two (dimensionful) parameters,
namely the temperature $T$ and the angular velocity $\Omega_H$. In the
following we will consider only di-rings for which \eqref{eqn:thecs} and
\eqref{eqn:conicalsingus} have been imposed.

Following \cite{Elvang:2007rd} (see also \cite{Evslin:2007fv}) a better
parametrization of the solution can be obtained by introducing an
overall length scale $L$,
\begin{equation}
 L^2=a_7-a_1\,,
\end{equation}
 and five dimensionless constants that specify the relative lengths of the rods:
\begin{equation}
 \kappa_i=\frac{a_{i+1}-a_1}{L^2}\,,\qquad i=1,\ldots, 5\,,
\end{equation}
so that $0<\kappa_1<\kappa_2<\kappa_3<\kappa_4<\kappa_5<1$. In
fig.~\ref{fig:diring} we depict the rod structure in terms of this
parametrization.  In addition, we find it useful to define 
\begin{equation}
 \bar c_1 = \frac{c_1}{L}=\sqrt{\frac{2\,\kappa_2\,\kappa_5}{\kappa_1\,\kappa_4}}\,,\qquad
\bar c_2=\frac{c_2}{L}=\sqrt{\frac{2(1-\kappa_3)(\kappa_3-\kappa_2)(\kappa_5-\kappa_3)}{(\kappa_3-\kappa_1)(\kappa_4-\kappa_3)}}\,.
\end{equation}

For the purpose of studying di-rings in thermodynamical equilibrium we will need some of the physical parameters of the individual black rings. We label the outer ring by ``1'' and the inner one by ``2''. The  physical parameters that we need are:

\paragraph{Horizon areas}
\begin{subequations}
\begin{align}
 A_H^{(1)}=&L^3\,4\,\sqrt2\,\pi^2\sqrt{\frac{(\kappa_2-\kappa_1)^3(\kappa_5-\kappa_1)}{\kappa_3^2(1-\kappa_1)^2(\kappa_4-\kappa_1)^2}}\\
&\times\sqrt{\kappa_1 (\kappa_3-\kappa_2) (1-\kappa_3) (\kappa_4-\kappa_3) (\kappa_5-\kappa_3)+\kappa_2 \kappa_4 \kappa_5 (\kappa_3-\kappa_1)-\bar c_1\bar c_2\kappa_1\kappa_4 (\kappa_3-\kappa_1) (\kappa_4-\kappa_3) }\,,\nonumber\\
 A_H^{(2)}=&L^3\,4\,\sqrt2\,\pi^2\sqrt{\frac{\kappa_5(1-\kappa_2)(1-\kappa_3)(\kappa_5-\kappa_1)(\kappa_5-\kappa_3)(\kappa_5-\kappa_4)^3}{(1-\kappa_1)^2(1-\kappa_4)^2(\kappa_5-\kappa_2)^2}}\,,
\end{align}
\end{subequations}
and the total horizon area is $A_H= A_H^{(1)}+ A_H^{(2)}$.

\paragraph{Temperatures}
\begin{subequations}
\begin{align}
T^{(1)}=&\,\,\frac{1}{2\,\pi L}\sqrt{\frac{\kappa_3^2(1-\kappa_1)^2(\kappa_4-\kappa_1)^2}{2(\kappa_2-\kappa_1)(\kappa_5-\kappa_1)}}\nonumber\\
&\times\big[\kappa_1 (\kappa_3-\kappa_2) (1-\kappa_3) (\kappa_4-\kappa_3) (\kappa_5-\kappa_3)+\kappa_2 \kappa_4 \kappa_5 (\kappa_3-\kappa_1)\nonumber\\
&\hspace{1cm}-\bar c_1\bar c_2\kappa_1\kappa_4 (\kappa_3-\kappa_1) (\kappa_4-\kappa_3) \big]^{-1/2}\\
T^{(2)}=&\,\,\frac{1}{2\,\pi L}\sqrt{\frac{(1-\kappa_1)^2 (1-\kappa_4)^2 (\kappa_5-\kappa_2)^2}{2\kappa_5(1-\kappa_2)(1-\kappa_3)(\kappa_5-\kappa_1)(\kappa_5-\kappa_3)(\kappa_5-\kappa_4)}}
\end{align}
\end{subequations}

\paragraph{Angular velocities}
\begin{subequations}
\begin{align}
&\Omega_H^{(1)}=\frac{1}{L}\,\frac{\bar c_1\,\kappa_1\,\kappa_3}{2\,\kappa_2\,\kappa_5-\bar c_1\,\bar c_2\,\kappa_1(\kappa_4-\kappa_3)}\,,\\
&\Omega_H^{(2)}=\frac{1}{2\,L}\,\frac{\bar c_1\,\kappa_1\,\kappa_4\,(1-\kappa_3)(\kappa_5-\kappa_3)+\bar c_2\,\kappa_5\,(\kappa_3-\kappa_1)(\kappa_4-\kappa_3)}{\kappa_3\,\kappa_5(1-\kappa_3)(\kappa_5-\kappa_3)}
\end{align}
\end{subequations}

\providecommand{\href}[2]{#2}\begingroup\raggedright
\endgroup


\end{document}